\documentclass[12pt]{article}

\newcommand{\be}{\begin{eqnarray}}
 \newcommand{\ee}{\end{eqnarray}}
 \newcommand{\nee}{\nonumber\end{eqnarray}}
 \newcommand{\nn}{\nonumber\\}

 \def\a{\alpha}

\usepackage{amsmath,amssymb,amsfonts,amsbsy}
\usepackage{graphicx}
\usepackage{wrapfig}

\begin{document}
\addcontentsline{toc}{subsection}{{Towards a model independent determination of fragmentation functions}\\
{\it B.B. Author-Speaker}}

\setcounter{section}{0}
\setcounter{subsection}{0}
\setcounter{equation}{0}
\setcounter{figure}{0}
\setcounter{footnote}{0}
\setcounter{table}{0}

\begin{center}
\textbf{TOWARDS A MODEL INDEPENDENT DETERMINATION OF FRAGMENTATION
FUNCTIONS}\footnote{talk given at the XIII workshop on High Energy Spin Physics, Dubna,
 Russia, 1-5 Sept. 2009, DSPIN-2009}

\vspace{5mm}

 \underline{E.  Christova}$^{\,1\,\dag}$,   E. Leader
$^{\,2\,\dag\dag}$ and S. Albino$^{\,3\,\dag\dag\dag}$,

\vspace{5mm}

\begin{small}
  (1) \emph{Institute for Nuclear Research and Nuclear Energy, Bulgarian Academy of Sciences, \\Sofia, Bulgaria} \\
  (2) \emph{Imperial College, London University, London, UK} \\
  (3) \emph{{II.} Institut f\"ur Theoretische Physik, Universit\"at Hamburg, Hamburg, Germany }\\
   \emph{E-mail:} $\dag$\emph{echristo@inrne.bas.bg}; $\dag\dag$\emph{e.leader@imperial.ac.uk}; $\dag\dag\dag$\emph{simon@mail.desy.de}
\end{small}
\end{center}

\vspace{0.0mm} 

\begin{abstract}
 We show that the difference cross sections in unpolarized SIDIS
$e+N\to e+h+X$ and $pp$ hadron production $p+p\to h+X$ determine,
uniquely and in any order in QCD, the
two FFs: $D_u^{h-\bar h}$ and $D_d^{h-\bar h}$, $h=\pi^\pm ,
K^\pm$. If both $K^\pm$ and $K_s^0$
are measured, then  $e^+e^-\to K+X$, $e+N\to e+K+X$ and $p+p\to
K+X$ yield independent measurements of
$(D_u-D_d)^{K^++K^-}$. In a combined fit to $K^\pm$ and $K^0_s$
production data from $e^+e^-$ collisions,  $(D_u-D_d)^{K^++K^-}$ is  obtained and compared to
 conventional parametrizations.
\end{abstract}


\section{Introduction}

Now, that
the new generation of high energy scattering
experiments with a final hadron $h$ detected are
taking place, it has
become clear that in order to obtain
the correct information about
quark-lepton interactions, not only
knowledge of the parton distribution functions (PDFs) 
are important, but a good knowledge of the fragmentation functions
(FFs) $D_i^h$, that determine the transition of  partons $i$ into
hadrons $h$, are equally important. The PDFs and the FFs are the two
basic ingredients that have to be correctly extracted from
experiment.  Here we
discuss the FFs.

The most direct way to determine the
FFs is the total cross section for
one-particle inclusive production in $e^+e^-$ annihilation:
\be
e^+e^-\to h+X,\qquad h=\pi^\pm,\,K^\pm,\,p/\bar p ...\label{e+e-}
\ee
However, these processes can
determine, in principle, only the combinations
\be
D_{u+\bar u}^h,\quad D_{d+\bar d}^h,\quad D_{s+\bar s}^h,
\quad D_g^{h+\bar h}\quad (D_{q+\bar q}^h=D_q^{h+\bar h}\equiv D_q^{h}+D_q^{\bar h}),
\ee
i.e.\ they
cannot distinguish the quark and anti-quark FFs. In order to achieve separate {\it phenomenological} determination of
$D_{q}^h$ and $D_{\bar q}^h$, the  one-hadron semi-inclusive
 processes  play an essential role:
  \be
l+N\,\to l+h+X \quad  {\rm and} \quad
p+p\,\to h+X.\label{pp}
\ee
The factorization theorem implies that the  FFs are universal, i.e. in (\ref{e+e-}) and
(\ref{pp}) the FFs are the same. However, in
(\ref{pp})  the hadron structure enters and when analyzing the data, usually different theoretical assumptions have to be made.

 Schematically the cross sections for (\ref{e+e-}) and (\ref{pp}) can be
written in the form:
\be
d\sigma^h_{e^+e^-}(z) &\simeq& \sum_c\hat\sigma_{e^+e^-}^c\otimes D_c^{h+\bar h}\\
d\sigma^h_N(x,z) &\simeq& \sum_{q,c} f_q(x)\otimes \hat\sigma_{lq}^c\otimes D_{c}^h\\
d\sigma^h_{pp}(x,z) &\simeq& \sum_{a,b,c} f_a\otimes
f_b\otimes\hat\sigma_{ab}^c\otimes D_c^h
\ee
where $\hat\sigma^c$
are the corresponding, perturbatively QCD calculable, parton-level
 cross sections for producing a parton $c$, $f_a$ are the
unpolarized PDFs.  At present several sets of FFs exist and
there is a significant disagreement among  some of the FFs.

 In this talk we present a model independent approach, developed in   \cite{FFs},
 which suggests that if instead of $d\sigma_N^h$ and $d\sigma_{pp}^h$ one works with
  the difference cross sections for producing  hadrons and producing their
  antiparticles, i.e. with data on
  $d\sigma_N^{h-\bar h}\equiv d\sigma_N^h-d\sigma_N^{\bar h}$ or
  $d\sigma_{pp}^{h-\bar h}\equiv d\sigma_{pp}^h-d\sigma_{pp}^{\bar h}$,  one obtains
   information about the non-singlet (NS) combinations $D_q^{h-\bar h}$. This is the complementary  to
   $D_q^{h+\bar h}$ quantity, measured in
   (\ref{e+e-}), that  would allow to determine $D_q$ and $D_{\bar q}$ without assumptions.

   Further this method is applied to
charged and neutral kaon $e^+e^-$-production data
to determine directly the non-singlet $(D_u-D_d)^{K^++K^-}$ and compare to conventional global fit analysis.

\section{Difference cross sections with $\pi^\pm$ and $K^\pm$}

From C-invariance of strong interactions it follows:
 \be
 D_g^{h^+-h^-}=0,\qquad
D_q^{h^+-h^-}=-D_{\bar q}^{h^+-h^-} \label{C-inv}
\ee
which, applied to (\ref{pp}), eliminates $D_g^{h^+-h^-}$ and $D_{\bar q}^{h^+-h^-}$ in the
difference cross sections:
\be
d\sigma_N^{h^+-h^-}=d\sigma_N^{h^+}-d\sigma_N^{h^-}\quad {\rm and} \quad
d\sigma_{pp}^{h^+-h^-}= d\sigma_{pp}^{h^+}-d\sigma_{pp}^{h^-}
\ee
 This implies that, in any order of QCD,  $d\sigma_N^{h^+-h^-}$ and
$d\sigma_{pp}^{h^+-h^-}$ are  expressed solely in terms of the NS combinations of the FFs.
In NLO  we have:
 \be
\hspace*{-0.6cm} d\sigma_p^{h^+-h^-}(x,z,Q^2) &=&
\frac{1}{9}\left[ 4u_V\otimes D_u+d_V\otimes  D_d+s_V\otimes
D_s\right]^{h^+-h^-}
 \otimes (1+\frac{\a_s}{2\pi} C_{qq})\label{p}\\
\hspace*{-.6cm}d\sigma_d^{h^+-h^-}(x,z,Q^2) &=&
\frac{1}{9}\left[(u_V+d_V)\otimes(4D_u+ D_d)+2s_V\otimes D_s\right]^{h^+-h^-}
\otimes (1+\frac{\a_s}{2\pi} C_{qq})\label{d}\\
\hspace*{-.6cm}E^h\frac{d\sigma_{pp}^{h^+-h^-}}{d^3P^h} &=&
\frac{1}{\pi}\int dx_a\,\int dx_b\,\int \frac{dz}{z}\times\nn
&&\times \sum_{q=u,d,s}\left[ L_q(x_b,t,u)q_V(x_a)
+L_q(x_a,u,t)q_V(x_b)\right]D_q^{h^+-h^-}(z) \label{ppdiff}
 \ee
where $u_V$ and $d_V$ are the valence quarks PDFs, $s_V=s-\bar
s$ and  $L_q$, given explicitely in \cite{FFs},  are  functions of the {\it known} quark densities $ q+\bar q$
and  partonic cross sections.

Common for the difference cross sections (\ref{p}) - (\ref{ppdiff}) is that
they all have the same structure:  1) only the non-singelts $D_q^{h-\bar h}$
 enter and 2) they enter in the combination $q_VD_q^{h-\bar h}$.
This implies that
  the contributions of $D_u^h$ and $D_d^h$ are enhanced by the large
 valence quark densities, while $D_s^h$ is suppressed by the small factor $(s-\bar s)$.
 Recently a strong bound on
$(s-\bar s)$ was obtained from neutrino experiments -- $\vert s-\bar s \vert \leq 0.025$~\cite{Soffer},
which  implies that
the contribution from  $D_s^h$ can be safely neglected. Thus, the $ep$, $ed$ and $pp$ semi-inclusive difference cross sections provide
three independent measurements  for the two unknown FFs $ D_u^{h^+-h^-}$ and $ D_d^{h^+-h^-}$.
 Note that the SIDIS  cross sections involve only $u_V$ and $d_V$, which are the best known parton densities,
  with  2\%-3\% accuracy at
$x\lesssim 0.7$.

Further information can be obtained specifying  the final hadrons.

1) If $h=\pi^\pm$ the difference cross sections will determine, without any assumptions,
$D_u^{\pi^+-\pi^-}$ and $D_d^{\pi^+-\pi^-}$ which would allow to test the usually made assumption
\be
D_u^{\pi^+-\pi^-} = -D_d^{\pi^+-\pi^-}.\label{SU2pions}
\ee
In \cite{DSS} it was suggested, for the first time, that this relation might be violated up to 10 \%.

2)  If $h=K^\pm$ the difference cross sections will determine, without any assumptions,
$D_u^{K^+-K^-}$ and $D_d^{K^+-K^-}$ which would allow to test the usually made assumption
\be
D_d^{K^+-K^-} = 0.\label{SU2K}
\ee

One can  formulate the above results also like this: If  relation
$D_u^{\pi^+-\pi^-} = -D_d^{\pi^+-\pi^-}$ (or $D_d^{K^+-K^-} = 0$) holds, then Eqs.~(\ref{p}), (\ref{d}) and (\ref{ppdiff}) for
$h=\pi^\pm$ (or $h=K^\pm$)
are expressed solely in terms of $D_u^{\pi^+-\pi^-}$ (or $D_u^{K^+-K^-}$) and thus look particularly simple.

\section{ Difference cross sections with $K^\pm$ and $K_s^0$ }

If in addition to the charged  $K^\pm$ also  neutral kaons $K_s^0=(K^0+\bar K^0)/\sqrt 2$
   are measured,  no new FFs are introduced into the cross-sections.
We  show that the combination
\be
\sigma ^{\cal K}\equiv \sigma ^{K^+}+\sigma ^{K^-}-2\sigma ^{K_s^0}
\ee
 \vspace*{-0.05cm}
in the considered three types of semi-inclusive processes (\ref{e+e-}) and (\ref{pp}):
 \be
&&e^++e^-\to K+X,\qquad \label{e+e-Ks}\\
&&e+N\to e +K+X,\qquad \qquad \label{SIDISKs}\\
&&p+p\to K+X,\qquad \label{ppKs}
 \ee
$K=K^\pm, K^0_s$, always measures only the  NS combination
$(D_u-D_d)^{K^++K^-}$.  This result relies  only on SU(2) invariance for the kaons,
 that relates $D_q^{K^0_s}$ to $D_q^{K^++K^-}$ and
does not involve {\it any} assumptions about PDFs or  FFs, it holds in any order in QCD.
As  $(D_u-D_d)^{K^++K^-}$, obtained in this way, is  model independent
 it would be interesting to compare it  to the existing parametrizations  from
$e^+e^-$ data,  obtained using  various assumptions.

\section{$(D_u-D_d)^{K^++K^-}$ from $e^+e^-$ kaon production}

The most precisely measured and theoretically calculated process is (\ref{e+e-Ks}). In NLO we
have~\cite{FFs}:
 \be
d\sigma_{e^+e^-}^{\cal K}(z,Q^2)=
\frac{8\pi\,\a_{em}^2}{s}\, (\hat e_u^2-\hat e_d^2)(1+\frac{\a_s}{2\pi} \, C_q\,\otimes \,)
 \,(D_u-D_d)^{K^++K^-}(z, Q^2).\label{e+e-K}
 \ee
where $\hat e_q^2(s)$ are the quark electro-weak charges. Using (\ref{e+e-K}) we determine~\cite{AC}
 $(D_u-D_d)^{K^++K^-}$ from the  available data on $K^\pm$ and $K^0_s$
production in $e^+e^-\to (\gamma ,Z)\to K+X,\quad K=K^\pm, K^0_s$ and
compare it to those obtained in global fit analysis.

 Our analysis has several advantages: it allows  for the first time
 to extract  $(D_u-D_d)^{K^++K^-}$  without any assumptions
about the  FFs (commonly used in  global fit
analysis) and without any
correlations to other FFs (and especially to $D_g^{K^\pm}$), it allows to  use
data at much lower values of $z$ than  in global fit
analysis (being a NS the
combination $\sigma^{\cal K}_{e^+e^-}$
does not contain the unresummed soft gluon
logarithms),  etc.

We have included the data on  $K^\pm$ and $K^0_s$ production from TASSO, HRS, MARKII,
TPC, TOPAZ, ALEPH, DELPHI, OPAL, SLD and CELLO collaborations in the energy intervals $\sqrt{s}=$
12 -- 14.8, 21.5 -- 22, 29 -- 35, 42.6 -- 44, 58, 91.2 and  183 -- 186 GeV.

In Fig.1 we present  $(D_u-D_d)^{K^++K^-}$ in NLO obtained in our approach
and from the global fits  of the DSS\cite{DSS}, HKNS\cite{HKNS} and AKK08\cite{AKK08}
sets.
  As seen from the Figure, at $z\gtrsim 0.4$ there is an  agreement
among the different FFs in  shape, but our FF is in general larger. The difference becomes
 more pronounced
  at $z \lesssim  0.4$.
 There could be different  reasons for this.
Most probably it is due either to inclusion of the small $z$-data  in our fit
or to the different assumptions in the global fit parametrizations.\\

{\bf\large  Acknowledgments}\\

\noindent
  The paper is supported by HEPTools EU network  MRTN-CT-2006-035505
and by Grant 288/2008  of Bulgarian National Science Foundation.

\begin{wrapfigure}[10]{R}{50mm}
  \centering 
  \vspace*{-8mm} 
  \includegraphics[width=50mm,,angle=-90]{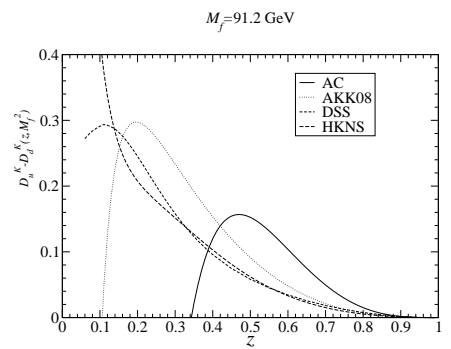}
  \caption{\footnotesize
$D_{u-d}^{K^++K^-}$ as obtained in this paper (labeled AC) and as obtained from the parametrizations of
DSS,  HKNS and AKK08. }
  \label{yourname_fig1}
\end{wrapfigure}

\end{document}